\documentclass[letterpaper]{article}
\usepackage{physics}
\usepackage{amsmath}
\usepackage{natbib,alifeconf} 
\usepackage{algorithm}
\usepackage{algorithmic}
\usepackage{diagbox}
\usepackage{hyperref}
\hypersetup{
    colorlinks=true,
    linkcolor=blue,
    filecolor=blue,      
    urlcolor=blue,
    citecolor=blue,
}

\title{A mechanism to promote social behaviour in household load balancing}
\author{Nathan A. Brooks$^{1,*}$, Simon T. Powers$^{2,3}$ \and James M. Borg$^{1}$ \\
\mbox{}\\
$^1$ School of Computing and Mathematics, Keele University, Staffordshire, UK, ST5 5BG\\
$^2$ School of Computing, Edinburgh Napier University, Edinburgh, UK, EH10 5DT \\
$^3$ Institute of Liberal Arts and Sciences, Keele University, Staffordshire, UK, ST5 5BG \\
$^*$ n.a.brooks@keele.ac.uk} % email of corresponding author
\begin{document}
\maketitle
\begin{abstract}
   Reducing the peak energy consumption of households is essential for the effective use of renewable energy sources, in order to ensure that as much household demand as possible can be met by renewable sources. This entails spreading out the use of high-powered appliances such as dishwashers and washing machines throughout the day. Traditional approaches to this problem have relied on differential pricing set by a centralised utility company. But this mechanism has not been effective in promoting widespread shifting of appliance usage. Here we consider an alternative decentralised mechanism, where agents receive an initial allocation of time-slots to use their appliances and can then exchange these with other agents. If agents are willing to be more flexible in the exchanges they accept, then overall satisfaction, in terms of the percentage of agents’ time-slot preferences that are satisfied, will increase. This requires a mechanism that can incentivise agents to be more flexible. Building on previous work, we show that a mechanism incorporating social capital –- the tracking of favours given and received –- can incentivise agents to act flexibly and give favours by accepting exchanges that do not immediately benefit them. We demonstrate that a mechanism that tracks favours increases the overall satisfaction of agents, and crucially allows social agents that give favours to outcompete selfish agents that do not under payoff-biased social learning. Thus, even completely self-interested agents are expected to learn to produce socially beneficial outcomes.          	 
\end{abstract}
\section{Introduction}
The UK government has committed to a legally binding target to reduce greenhouse gas emissions by 80\% by 2050. This requires a shift to renewable energy sources, such as solar panels and wind turbines. However, integrating renewable energy sources into a centralised and monolithic `national grid' is difficult because their output inherently depends on weather conditions. As such, they cannot simply be `switched on and off' to meet demand in the way that coal, gas, and nuclear power stations can be to match supply and demand. Governments and energy providers have recognised that this problem of load balancing -- matching supply and demand -- is easier to solve on a local scale. Consequently, they are supporting the development of `community energy systems', where a community (e.g. a town or a small island) owns and manages its own renewable energy sources \citep{Walker:2008:a}.

The shift towards community energy systems means that communities now become involved in some of the tasks that were previously handled by a centralised national grid. In particular, they now become involved in the balancing of supply and demand. A key problem here is how to reduce \emph{peak} demand, i.e. the maximal amount of electricity that is demanded at any one moment in time. If a community's peak demand is too high, then it is unlikely that it will be able to be met by the community's renewable energy sources, and so the community is likely to have to resort to buying in electricity from non-renewable sources. 

%But if the demand could be spread out more evenly throughout the day, then all of it may be met from their renewable sources.

The traditional approach to reducing peak demand is differential pricing set by a central utility company. Simply put, households are incentivised to run their appliances at times of low demand through lower pricing at these times \citep{Stern:1986:a,Dutta:2017:a}. Traditionally, this has involved utility companies offering cheaper electricity overnight. Could a community energy system use the same mechanism? Potentially, however, variable pricing inherently discriminates against more vulnerable households on lower incomes \citep{Simmons:2007:a}. Then there is the question of how prices should be set and who should set them? People are unlikely to take part in such a scheme unless they perceive that they are being treated fairly. 

%
%however solar panels, for example, require load to be spread out more evenly throughout daylight hours

To address these issues, we consider an alternative mechanism for load balancing in a community energy system, which is not based on pricing set by some centralised authority. We assume that each household will have preferred time-slots for when they would like to run high-powered appliances such as washing machines, dishwashers and electric heating. The aim is then to allocate actual time-slots to each household for when they run their appliances. On the one hand, this is a classic multi-objective optimisation problem of reducing peak consumption (the maximum amount of energy demanded in a time-slot) while satisfying each households' preferences as far as possible. On the other hand, issues of fairness are central. If households are to be motivated to use the mechanism then they will need to perceive the resulting \emph{allocation} of time-slots to households as being fair (distributive justice, \citealt{Rescher:1966:a}). Furthermore, households will need to be able to understand why some of their time-slot preferences have not been met, and why the preferences of some other households may have been met instead. In other words, they need to perceive the allocation \emph{procedure} as treating them fairly (procedural justice, \citealt{Hollander-Blumoff:2008:a}).

\citet{Petruzzi:2013:a} propose a mechanism inspired by the building of social capital between agents (households, or software agents representing them). In their mechanism, agents are initially allocated time-slots at random, but can then propose exchanges of time-slots with other agents. Agents have two possible strategies. Selfish agents only accept exchanges that provide them with one of their preferred time-slots. Social agents, on the other hand, accept not only these beneficial exchanges but also accept an exchange request if they owe a favour to another agent (provided the exchange will not cause them to lose one of their preferred time-slots).. An agent owes a favour to another if the other agent has previously accepted an exchange request from them.
\citet{Petruzzi:2013:a} showed that under this mechanism, a group where every agent was social had on average more of their time-slot preferences satisfied than a group where every agent was selfish. They construed the recording of favours given and received, and the acting upon this by social agents, as the accumulation of a form of electronic social capital \citep{Putnam:1994:a,Petruzzi:2014:a}. Presumably, this would be intuitive for households to understand.

%and would fit well with physical scenarios where a community already has social capital, such as contributing to the upkeep of communal living spaces or collecting goods for neighbours.

However, two important questions arise from this work. First, to what extent is social capital a necessary part of the mechanism? Would social agents that always accept exchanges that do not cause them to lose a preferred time-slot also reach outcomes with high average satisfaction, without the need to track favours given and received? Second, should we expect self-interested agents to adopt the selfish or the social strategy? We address both of these questions in this paper. To address the first, we re-examine the \citet{Petruzzi:2013:a} model by allowing social capital to be turned off. To address the second, we consider mixed populations of selfish and social agents that change their strategy according to payoff-biased social learning \citep{Boyd:1985:a}. We find that while a mechanism without social capital allows a pure population of social agents to perform better than a pure population of selfish agents, when agents can change their strategy through social learning then social capital is necessary for social agents to outcompete selfish agents. Our results demonstrate that a time-slot allocation mechanism based on social capital can reduce peak electricity consumption, promote social behaviour, and lead to outcomes where the average satisfaction of households is high, even when agents are entirely self-interested.   

%in the discussion, talk about the benefits of decentralisation, i.e. suitable for use in a community energy system. Also talk about agents could be sanctioned for not using their allocated time-slots or could be given cheaper energy in their allocated time-slots.

%it is not a social dilemma but a coordination problem. Don't want selfish as won't be able to solve the optimisation problem so quickly in a dynamically changing environment.  Social capital eradicates selfish behaviour. Look at adding opportunity cost for doing an exchange in journal paper. This would make it into a social dilemma.

\section{The Energy Exchange Simulation}
%The Energy Exchange Simulation has been developed in order to better understand how social capital, in the form of trust, can influence direct interactions between agents in pairwise situations, and how this in turn can impact on the success of a population in solving a multi-objective optimisation problem.

The \citet{Petruzzi:2013:a} model was built to represent a smart energy network consisting of 96 individual agents. Each day agents request four hour-long time-slots in which they require electricity. All requests are for 1KWh of energy, and there can never be more than 16 agents using the same time-slot, as this is considered the peak capacity of the system. Time-slots are initially allocated at random at the start of the day, so few agents are likely to have their allocation match all of their requested time-slots. Because of this, after the initial allocation agents can partake in pairwise exchanges where one agent requests to swap one of its time-slots with a second agent, and the second agent decides whether or not to fulfil the request. We define an agent's \emph{satisfaction} as the proportion of its time-slot preferences that have been satisfied, and track the mean value of this as a measure of how well the mechanism is satisfying the agents' preferences.

%The \citet{Petruzzi:2013:a} model was built to represent a smart energy network consisting of 96 individual agents who have varying preferences for which time-slots they wish to use high powered appliances.

Agents can follow either social or selfish strategies, which impacts how they react to incoming requests for exchanges. Selfish agents will only accept exchanges that are in their immediate interest. This means that selfish agents need to be offered a time-slot that they have initially requested and do not already have in order for them to agree to the exchange. Social agents also agree to these mutually beneficial exchanges. However, they also make decisions based on social capital, in the form of repaying previous favours given to them by other agents. Specifically, when a social agent's request is accepted, they record it as a favour given to them. When a social agent receives a request from another agent who previously gave them a favour, they will accept the request if it is not detrimental to their own satisfaction and record that the favour has been repaid. This improves the satisfaction of the other agent while earning themselves more social capital. This leads to a system of social agents earning and repaying favours among one another, increasing the number of accepted exchange requests.

Exchanges begin on a day once each agent has received their initial allocation and decided which of these time-slots they wish to keep. They then anonymously advertise slots that they have been allocated but do not want to an `advertising board'. Several exchange rounds then take place during the day, where the number of rounds is a parameter of the model that sets the maximum number of exchanges an agent can engage in per day. In each exchange round, agents can request a time-slot from the board, so long as they have not already received a request from another agent during that round. Agents accept or refuse requests based on their strategy as described above. Only social capital, i.e. social agents' memory of favours, remains between days. 

We expand on the original \citet{Petruzzi:2013:a} model by introducing social learning, allowing agents to change from selfish to social or vice versa (note that both social and selfish agents undergo `social' learning, which we refer to simply as `learning' from now on to avoid confusion). This works as follows. At the end of each day, a percentage of the agents are randomly selected to undergo learning. Each agent performing learning observes a randomly selected second agent. If the observed agent has a higher satisfaction than the agent in question, then the first agent will copy the strategy of the observed agent with a probability proportionate to the difference between the two agents' satisfactions. Learning is thus payoff-biased \citep{Boyd:1985:a}, with strategies giving higher individual satisfaction more likely to spread in the population. Agents that move from a social strategy to the selfish strategy retain their accumulated social capital. Pseudocode for the simulation procedure is given in Algorithm~\ref{alg:Algorithm}\footnote{Source code is available at \url{https://github.com/NathanABrooks/ResourceExchangeArena}}.

\begin{algorithm}
    \begin{algorithmic}[1]
     \STATE $d \leftarrow $ current day
     \STATE $e \leftarrow $ current exchange\_round
     \STATE $A \leftarrow $ set of $a$ agents
     \STATE $L \leftarrow $ number of agents $a$ undergoing learning
    \FOR {$d$ = 1 to MAX\_DAYS}
     \FOR {each $a \in  A$ }
      \STATE $a$.receive\_random\_allocation()
     \ENDFOR
     \FOR {$e$ = 1 to MAX\_EXCHANGES}
      \STATE $V \leftarrow $ set of $v$ adverts
      \FOR {each $a \in  A$ }
       \STATE $v \leftarrow a$.determine\_unwanted\_time\_slots()
       \STATE $V$.list\_advert($v$)
      \ENDFOR
      \FOR {each $a \in  A$ }
       \IF {$a$.received\_request() == true }
        \STATE go to next agent
        \ENDIF
       \IF {$a$.satisfaction() == 1 }
        \STATE go to next agent
       \ELSE
        \STATE $r \leftarrow a$.identify\_beneficial\_exchange($V$)
        \STATE $a$.request\_exchange($r$)
       \ENDIF
      \ENDFOR
      \FOR {each $a \in  A$ }
       \IF {$a$.received\_request() == true }
        \STATE $a$.accept\_exchange\_if\_approved()
       \ENDIF
      \ENDFOR
      \FOR {each $a \in A$ }
       \IF {$a$.made\_exchange() \AND $a$.agent\_type == Social}
        \STATE $a$.update\_social\_capital()
       \ENDIF
      \ENDFOR
     \ENDFOR
     \STATE $d_l \leftarrow $ 0
     \WHILE{$d_l$ \textless \ $L$}
      \STATE {$a_1 \leftarrow $ random agent that hasn't considered changing strategy today}
      \STATE {$a_2 \leftarrow $ random agent to observe}
      \IF {$a_1$.satisfaction() \textless \ $a_2$.satisfaction() }
       \STATE {$x \leftarrow $ random value between 0 and 1}
       \IF{($a_2$.satisfaction() - $a_1$.satisfaction()) \textgreater \ $x$}
        \STATE $a_1$.copy\_strategy($a_2$)
       \ENDIF
      \ENDIF
      \STATE $d_l \leftarrow d_l$ + 1
     \ENDWHILE
    \ENDFOR
    \end{algorithmic}
    \caption{The Energy Exchange Simulation.}
    \label{alg:Algorithm}
\end{algorithm}
% I think to have this be a probability (e raised to the power of the difference) would be better, as at the moment the model can get "stuck" by this deterministic rule. We can do that for the journal version, though! Also look at adding a cost to exchanges to make it a social dilemma.

We also consider, as a counterpoint to social capital, simulations where social capital is not recorded by agents, resulting in social agents accepting any non-detrimental exchange.  We refer to the social agents under this mechanism as \emph{social without social capital}. All simulation results are averaged over 50 runs for each variation of the simulation parameters. The parameters that we vary are the number of exchange rounds per day, and the number of agents undergoing social learning at the end of each day (the learning rate). We hold the other parameters constant across simulations with the values given in Table~\ref{tab:Params}.

\begin{table}[h]
    \centering
    
    \begin{tabular}{ |l|l| } 
    \hline
    \textbf{Parameter} & \textbf{Value} \\
    \hline
    Population size & 96 \\
    Number of days & 500 \\
    Time-slots per day & 24 \\
    Slots selected by each agent & 4 \\
    Maximum agents per time-slot & 16 \\
    Simulation runs & 50 \\
    \hline
    \end{tabular}
    \caption{Constant parameter values.}
    \label{tab:Params}
\end{table}

\begin{figure}
    \centering
    \includegraphics[width=\linewidth]{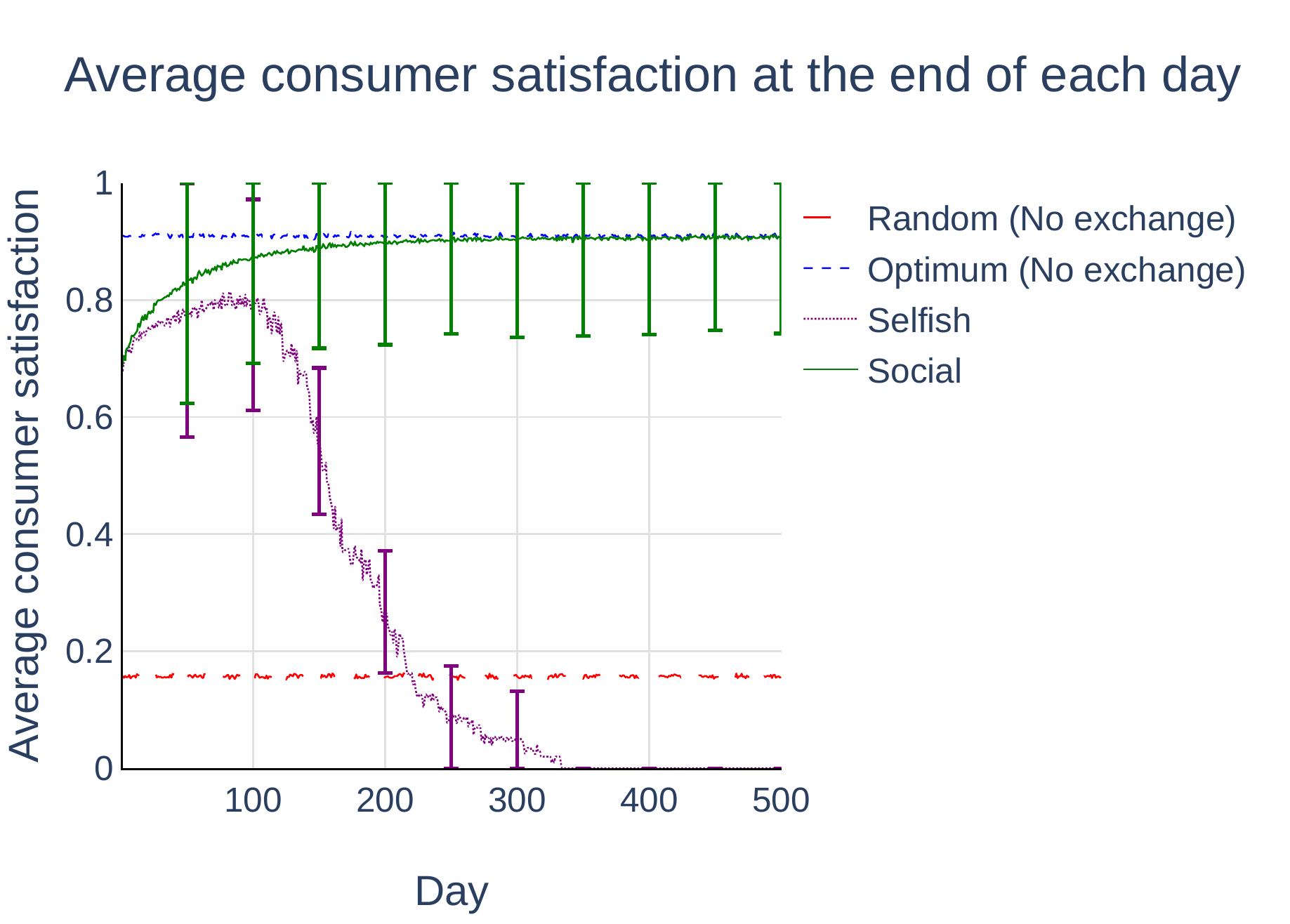}
    \caption{Average satisfaction per day for a typical run of the simulation. The run shown here includes social capital, uses a learning rate of 50\%, and 100 exchange rounds per day. The optimum result is calculated as the proportion of requested time-slots that exist within the simulation and could therefore be allocated to an agent who requested them.}
    \label{fig:Typical}
\end{figure}

Results from an illustrative run of the Energy Exchange Simulation can be seen in Figure~\ref{fig:Typical}, showing how agent satisfaction changes over time. 

%In this simulation, the population started out with 50\% selfish agents and 50\% social agents. 

\section{Results and Analysis}
To answer the research questions raised in the introduction, our analysis of the simulation proceeds as follows. We first consider populations where every agent uses the same strategy, and there is no learning. This allows us to determine whether a pure population of social agents using the mechanism without social capital can do as well as a pure population of agents that do track social capital. We then go on to consider mixed populations where both selfish and social agents are present and can switch their strategies through learning. This allows us to determine the conditions under which social agents can outcompete selfish agents, and the extent to which this is affected by whether social capital is tracked.
\subsection{Single Strategy Populations}
\begin{figure*}
    \centering
    \includegraphics[width=0.75\textwidth]{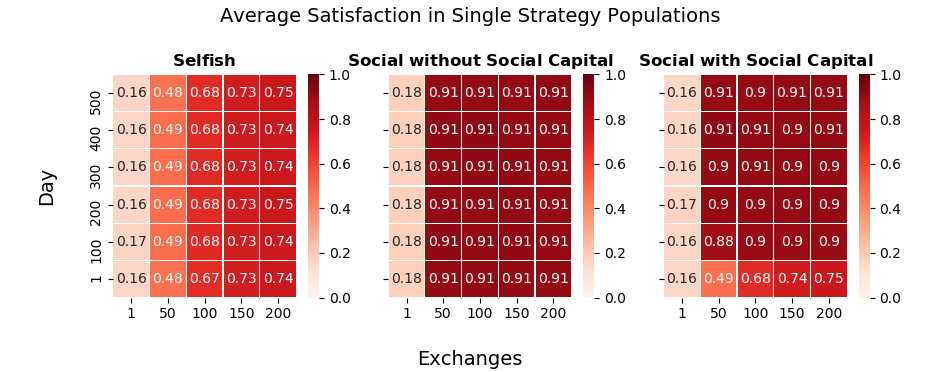}
    \caption{Average satisfaction of single strategy populations. For all plots the $y$-axis shows the simulated day satisfaction was measured on, and the $x$-axis shows the number of exchanges an agent is permitted to engage in per day. (Left) Average satisfaction for populations of selfish agents. (Middle) Average satisfaction for populations of social agents without social capital. (Right) Average satisfaction for populations of social agents with social capital.}
    \label{fig:onestrat}
\end{figure*}

To establish a baseline for the performance of each of the available strategies explored here (selfish, and social with and without social capital), each is first run in isolation of the others, i.e. as a pure population of the strategy. Average (mean) satisfaction heatmaps for these simulations, across all parameters, can be seen in Figure \ref{fig:onestrat}. Optimal performance in these simulations would result in an average satisfaction of approximately 0.91 (as seen in Figure \ref{fig:Typical}). From Figure \ref{fig:onestrat} we can see that social populations with social capital achieve this optimal performance over the vast majority of parameters settings. Selfish populations, on the other hand, consistently fall short of optimal satisfaction, but do show improvement as the number of rounds of exchanges on a day increases. No improvement is seen as the number of days the simulation is run for is increased, which follows from the fact that the only state carried over between days is social capital, with selfish agents do not use. Social populations without the social capital mechanism similarly exhibit no improvement as the number of days increases for the same reason. But crucially, they reach optimal performance on the first day, if afforded enough exchanges. The inclusion of social capital means that agents require more days to reach (near) optimal performance, with this only being achieved after 100 days of social capital accumulation. 

Overall, it is clear from Figure \ref{fig:onestrat} that satisfaction is dependent on the number of exchanges available to agents, and that social agents are able to break of out sub-optimal states more easily than selfish agents. This is through social agents accepting exchanges that are neutral to their satisfaction but that increase the satisfaction of another agent. However, the tracking and consideration of social capital slows this process down compared to the mechanism without social capital. This is because with tracking of social capital, social agents will only accept neutral exchanges if they owe a favour to the other agent, and so they behave like selfish agents until they begin to owe favours from previous days. 
\subsection{Mixed Populations without Social Capital}
\begin{figure*}[ht!]
    \centering
    \includegraphics[width=0.75\textwidth, height=0.5\textheight]{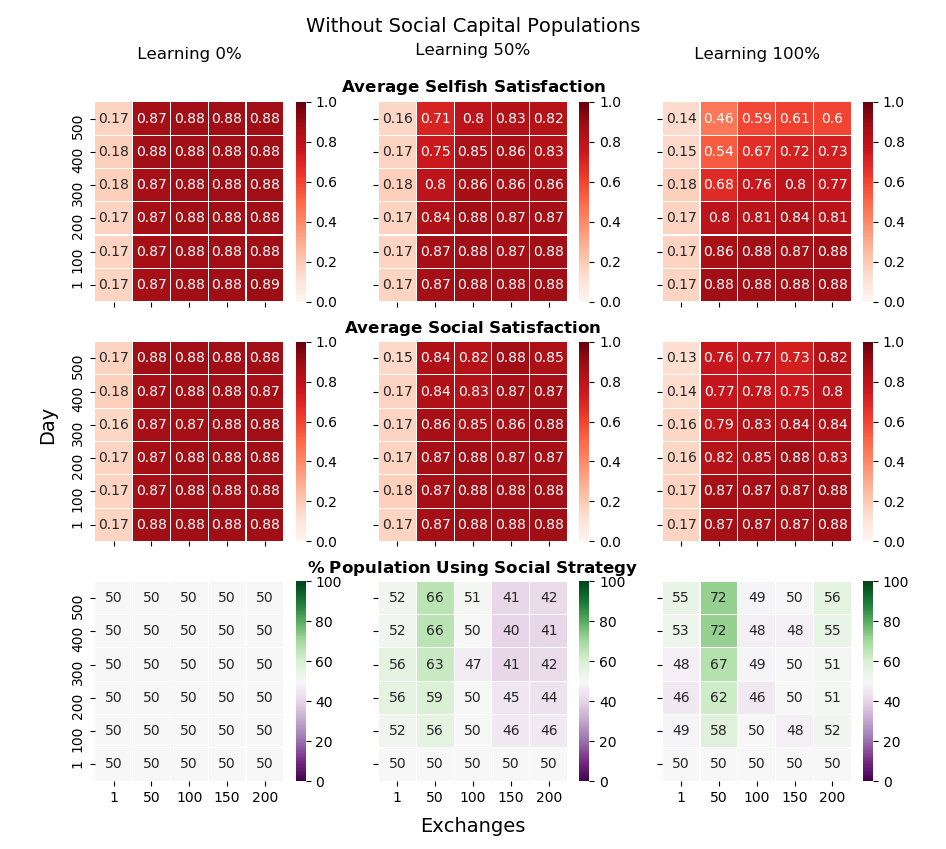}
    \caption{Average satisfaction of selfish and social agents in mixed populations without social capital. For all plots the $y$-axis shows the simulated day satisfaction was measured on, and the $x$-axis shows the number of exchanges an agent is permitted to engage in per day. (Top Row) Average satisfaction of selfish agents. (Middle Row) Average satisfaction of social agents. (Bottom Row) Proportion of population using the social strategy; green indicates a greater proportion of social agents; purple indicates a greater proportion of selfish agents.}
    \label{fig:WSCAll}
\end{figure*}

Results for populations which combine selfish and social agents, but without access to social capital (resulting in social agents always accepting non-detrimental exchanges) can be seen in Figure \ref{fig:WSCAll}. In these simulations, populations start with an equal number of social and selfish agents, with learning taking place at the end of each day, affecting the overall ratio of social to selfish agents. Different learning rates are also tested; 0\%, resulting in all agents retaining the strategy they were initialised with, 50\%, permitting half of the agents (at random) to undergo learning, and 100\%, resulting in all agents engaging in learning per day. Figure~\ref{fig:WSCAll} (Bottom Row) shows the change in strategy distribution across all parameters and learning rates. Within the maximum number of days available in these simulations neither the social nor selfish strategy is eradicated from the population (which is possible with the payoff-biased learning in Algorithm~1). In fact, it is often the case that close to a 50:50 distribution of strategies is retained, indicating no significant change due to learning. We do find that the social strategy begins to dominate the population when 50 exchanges per day are allowed, the simulation has run for over half the maximum number of days, and learning is implemented. But when the learning rate is set to 50\%, and the number of exchanges exceeds 150, we see the population distributions beginning to swing in the favour of the selfish strategy. Increasing the learning rate to 100\% does stop the selfish strategy gaining prominence under these conditions but does not even get close to eradicating the selfish strategy.

Given the (near) optimal performance demonstrated by social agents (without social capital) observed in Figure \ref{fig:onestrat}, where the social strategy existed in isolation, it is clear from Figure \ref{fig:WSCAll} (Middle Row) that the inclusion of the selfish strategy has a detrimental effect of the satisfaction of social agents, whilst improving the satisfaction of selfish agents (Figure \ref{fig:WSCAll} (Top Row)). The only time the social strategy achieves an advantage over the selfish strategy is when exchanges are set at 50 per day -- this result appears to be more a result of 50 exchanges not being sufficient to allow the selfish strategy to achieve its best performance, whilst being sufficient for the social strategy to do so. The negative consequences of the continued use of the selfish strategy to overall satisfaction of the populations, compared to when social strategies are applied in isolation, are clearly apparent. Whilst the social strategy is necessary for selfish agents to improve their satisfaction, through exploitation of the social agents' willingness to exchange, it is to the detriment of the overall satisfaction of the population. It is therefore in the interests of the populations to remove the selfish strategy. In effect, without social capital selfish agents are able to parasitise social agents to the detriment of the population as a whole.
\subsection{Mixed Populations with Social Capital}
\begin{figure*}[ht!]
    \centering
    \includegraphics[width=0.75\textwidth, height=0.5\textheight]{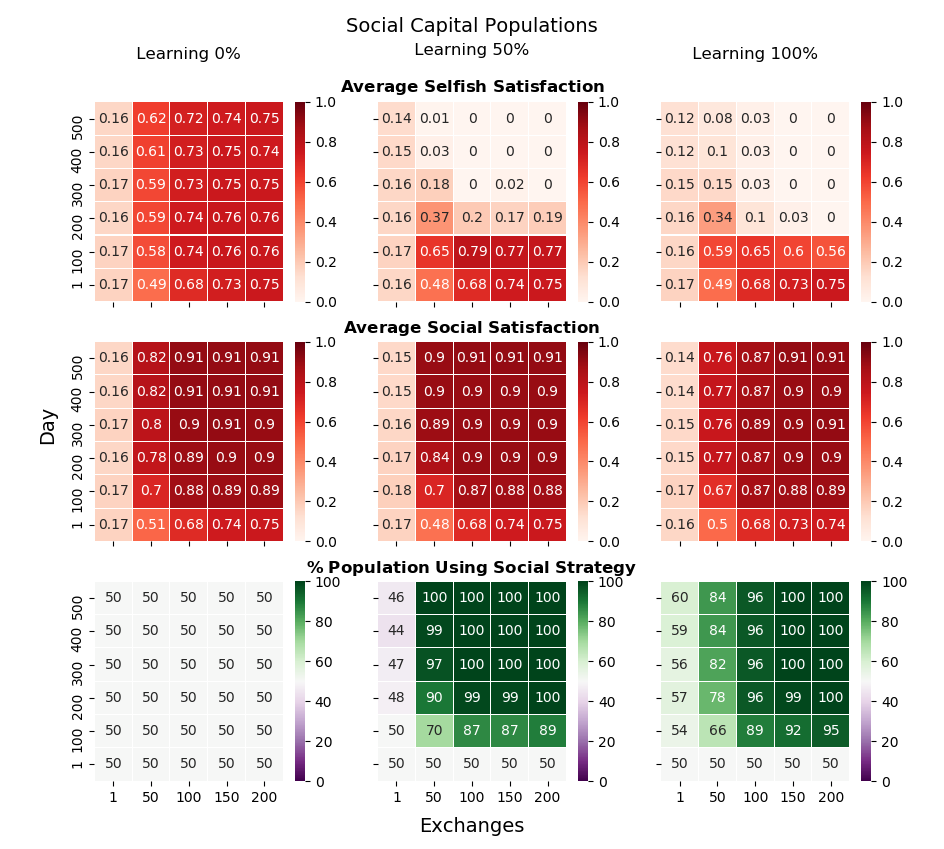}
    \caption{Average satisfaction of selfish and social agents in mixed populations with social capital. For all plots the $y$-axis shows the simulated day satisfaction was measured on, and the $x$-axis shows the number of exchanges an agent is permitted to engage in per day. (Top Row) Average satisfaction of selfish agents. (Middle Row) Average satisfaction of social agents. (Bottom Row) Proportion of population using the social strategy; green indicates a greater proportion of social agents; purple indicates a greater proportion of selfish agents.}
    \label{fig:SCAll}
\end{figure*}

% Following on from the previous section focus on the eradication of the selfish strategy, and the satisfaction improvement that is obtained as a result. But, do mention that this takes time and exchanges, but more exchanges requires fewer days, and more days requires fewer exchanges.
As in Figure \ref{fig:onestrat}, populations using the social strategy whilst keeping track of social capital exhibit a slower progression toward optimal satisfaction, as they only start accepting neutral exchanges as social capital accumulates over the days. This is also the case when the selfish strategy is included in a combined population (Figure \ref{fig:SCAll}). However, unlike mixed populations where social capital is not tracked (see Figure \ref{fig:WSCAll}), populations tracking and using social capital are able to achieve optimal (or near optimal) satisfaction when afforded enough exchanges, and the simulation is ran for over 100 days. As the learning rate is increased to 50\% the number of exchanges required to achieve near optimal average satisfaction scores amongst social agents drops, though it is interesting to note that increasing the learning rate again to 100\% does cause a drop in satisfaction when fewer exchanges per day are permitted. 

Social capital thus has the effect of slowing down optimisation, but with the benefit of hindering the ability of selfish agents to gain any traction in the population. We can see that when learning is set to 0\% selfish agents, despite making up half of the population, struggle achieve the same kinds of satisfaction scores as previously seen when social capital is not available (compare Figure~\ref{fig:WSCAll} (Left) to Figure~\ref{fig:SCAll} (Left)). As these selfish agents cannot accumulate social capital, desired time-slots held by other agents are harder to obtain than when social capital is not tracked. This enables social agents to build up exchange networks using social capital that cannot be invaded by selfish agents. When learning is introduced, the inclusion of social capital effectively eradicates the selfish strategy from the population. As social agents are on average more satisfied than selfish ones, with this effect becoming more apparent as the simulation goes on for more days or more exchanges are permitted, selfish agents increasingly switch strategies. Conversely, with no opportunity for selfish agents to parasitise the population there is little gradient for social agents to switch to the selfish strategy. We can see in Figure \ref{fig:SCAll} (Bottom) that inclusion of learning and social capital results in the entire population adopting the social strategy. Increasing learning to 100\% does slow the removal of the selfish strategy down, as it introduces the possibility of social agents switching to the selfish strategy in cases where some selfish agents have achieved high satisfaction by luck alone (i.e. being randomly allocated all of their preferred time slots), but generally learning combined with social capital has the effects of removing the selfish strategy form the population, leading to near optimal satisfaction scores when provided with enough exchanges and enough simulated days. 

\subsection{The Population-level Effect of Social Capital}
\begin{figure*}[ht!]
    \centering
    \includegraphics[width=0.75\textwidth]{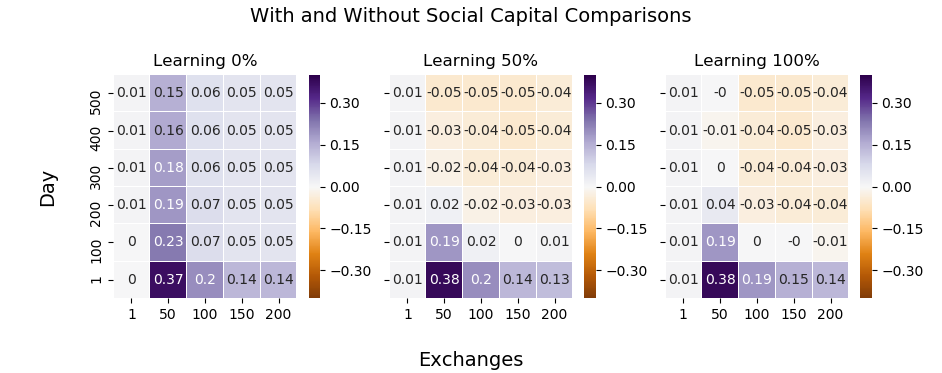}
    \caption{Average difference in satisfaction between all agents in populations with and without social capital. In all rows purple indicates higher satisfaction in populations without social capital, orange indicates higher satisfaction in populations with social capital. For all plots the $y$-axis shows the simulated day satisfaction was measured on, and the $x$-axis shows the number of exchanges an agent is permitted to engage in per day.}
    \label{fig:Comp}
\end{figure*}

%Re-stress the benefit of social capital - removal of selfish strategy and improvement of population level satisfaction.
Despite the introduction of social capital removing the selfish strategy, and achieving near optimal satisfaction across a number of parameter settings, it is still the case the overall population satisfactions (averaged over both selfish and social agents) are similar when comparing populations with and without social capital (Figure \ref{fig:Comp}). Due to social capital slowing down the rate at which social agents undergo exchanges, when only a low number of exchanges per day are available, or the simulations are stopped after a few days, populations without social capital actually outperform populations with social capital, despite selfish agents being largely present in populations without social capital. Once learning is introduced, the number of exchanges per day are increased, and the simulation is permitted to run for extended amount of time, we do begin to see social capital populations outperforming those without social capital, though only by a small margin. Taking just the raw satisfaction results for the 500\textsuperscript{th} day, we do see a significant difference between the satisfaction of social capital and non social capital populations over most parameter settings. Using a Mann-Whitney U test, we observe $p$-values where $p<0.01$ over most parameter settings (see Table~\ref{tab:stats}), the exceptions being when learning is 50\% and exchanges are set to 1, and when learning is 100\% and exchanges are low (1 or 50). These results indicate that by the 500\textsuperscript{th} day of the simulation, social capital populations where learning is enabled (at 50\% or above) are significantly more satisfied than non social capital populations. 

%One could also speculate that the removal of the selfish strategy as a viable approach to optimising satisfaction would have a secondary benefit in the real world, by increasing the sense of distributive and procedure justice amongst users.

\begin{table}[ht]
    \centering
    \begin{tabular}{|c||*{3}{c|}}\hline
    \multicolumn{4}{|c|}{Mann-Whitney U: $p<0.01$} \\
    \hline
    \backslashbox{Exchanges}{Learning}
    &\makebox{0\%}&\makebox{50\%}&\makebox{100\%}\\\hline\hline
    1 &True &False &False\\\hline
    50 &True &True &False\\\hline
    100 &True &True &True\\\hline
    150 &True &True &True\\\hline
    200 &True &True &True\\\hline
\end{tabular}
\caption{Mann-Whitney U test $p$-values when comparing satisfaction of agents in populations with and without social capital. Test conducted for 500\textsuperscript{th} day only. $p<0.01 = True$, $p\geq0.01 = False$.}
    \label{tab:stats}
\end{table}
\section{Discussion}
We have extended the \citet{Petruzzi:2013:a} model to answer two research questions. Our first question was, to what extent is the tracking of social capital a necessary feature of the exchange mechanism? We demonstrated that if social agents exist in isolation in a pure population, then they end up more satisfied when social capital is \emph{not} tracked, for cases where the number of days that the simulation is run for is low. When the number of days is increased, they perform similar to when social capital is tracked. Thus, in pure populations with no possibility of selfish agents being introduced then tracking of social capital is not beneficial. Moreover, tracking social capital is detrimental under some conditions since it requires more days before social agents start accepting neutral exchanges, and hence more days before they escape from sub-optimal initial allocations.

However, in a real system we must allow for the possibility that agents may act selfishly, or may learn to do so over time. To account for this, we have considered mixed populations where both social and selfish agents are present and can exchange with each other. In this case, social agents are able to achieve (near to) optimal satisfaction when social capital is tracked, whereas without social capital they do less well. This effect becomes more pronounced once agents are permitted to adjust their strategy by payoff-biased learning. This is because without social capital, selfish agents are able to persist in the population under payoff-biased learning. Their persistence is detrimental both to social agents and to the population as a whole, since by not accepting neutral exchanges they can prevent the agents escaping from sub-optimal allocations. Conversely, with social capital selfish agents are effectively purged from the population under payoff-biased learning, which has a significant positive effect on the satisfaction of the collective population.

The results with payoff-biased (social) learning also answer our second research question: should we expect self-interested agents to adopt the selfish or the social strategy? This is because under payoff-biased learning, agents will only change their strategy if they see another agent is doing better than itself under another strategy. Thus, our results suggest that self-interested agents should adopt the social strategy if social capital is tracked, but without social capital there is no pressure for them to do so.

Recent research has shown that a diverse range of agent-based mechanisms can be effective at managing community energy systems. Allowing agents to form self-organised clusters working to optimise their collective performance has shown to be an effective approach with larger population sizes \citep{Cauvsevic:2017:a}. There are also more complex algorithms that have the potential to be highly effective at managing decentralised heating systems\citep{Kolen:2017:a, Dengiz:2020:a}. Our work differs in that it is inherently human facing. A real world implementation of our system could easily operate in a socio-technical manner in which individuals can take over from the virtual agent representing them, setting their own preferences for time-slots and making decisions on whether or not to accept requested exchanges. Utilising a system based on social capital represented as `favours' would also be easy for the average user to understand facilitating procedural justice.

In conclusion, we have demonstrated a decentralised mechanism for household load balancing that is effective at satisfying agents' preferences. The benefits of a decentralised mechanism are that it is inherently scalable as more agents are introduced \citep{Petruzzi:2013:a}, and helps to promote privacy and trust by not requiring households to submit their time-slot preferences to a centralised authority. In a real implementation, the exchanges may be performed by software agents running on home gateways installed in households. This could involve various levels of user engagement with the exchange process. The mechanism could also be used alongside differential pricing -- households could be given a cheaper rate in their allocated time-slots. Future work should empirically investigate how users perceive distributive and procedural justice both with and without tracking social capital \citep{Powers:2019:a}.    

\section{Acknowledgements}
This work has been supported by a fellowship to Simon T. Powers from the Keele University Institute of Liberal Arts and Sciences.
\bibliographystyle{apalike}
\bibliography{energy.bib}
\end{document}